\documentclass[a4paper,11pt]{article}
\usepackage{graphicx,amssymb,bm,latexsym,color,epsf}
\makeatletter
\@addtoreset{equation}{section}

\makeatother
\newcommand{\be}{\begin{equation}}
\newcommand{\ee}{\end{equation}}
\newcommand{\bea}{\begin{eqnarray}}
\newcommand{\eea}{\end{eqnarray}}
\newcommand{\ena}{\end{eqnarray}}
\newcommand{\vs}[1]{\vspace{#1 mm}}

\newcommand{\cR}{{\cal R}}

\newcommand{\bR}{\bar R}
\newcommand{\bg}{\bar g}
\newcommand{\bh}{\bar h}
\newcommand{\bDelta}{\bar\Delta}

\newcommand{\bnabla}{\bar\nabla}
\newcommand{\tr}{{\rm tr}}
\newcommand{\Tr}{{\rm Tr}}

\newcommand{\hperp}{h^\perp}
\newcommand{\jgh}{J}
\newcommand{\jaux}{K}
\newcommand{\Lie}{{\cal L}}

\def\mg{\mathbf{g}}
\def\mbg{\bar\mathbf{g}}
\def\mh{\mathbf{h}}
\def\mX{\mathbf{X}}
\def\mY{\mathbf{Y}}

\begin{document}

\vs{10}
\begin{center}
{\Large\bf The background scale Ward identity
\\ in quantum gravity}
\vs{15}

{\large
Roberto Percacci\footnote{e-mail address: percacci@sissa.it}$^{,b,c}$
and Gian Paolo Vacca\footnote{e-mail address: vacca@bo.infn.it}$^{,d}$
} \\
\vs{10}

$^b${\em International School for Advanced Studies, via Bonomea 265, I-34136 Trieste, Italy}

$^c${\em INFN, Sezione di Trieste, Italy}

$^d${\em INFN, Sezione di Bologna, via Irnerio 46, I-40126 Bologna, Italy}

\vs{15}
{\bf Abstract}
\end{center}
We show that with suitable choices of parametrization, gauge 
fixing and cutoff, 
the anomalous variation of  the effective action 
under global rescalings of the background metric is identical to
the derivative with respect to the cutoff,
{\it i.e.} to the beta functional, as defined by the exact RG equation.
The Ward identity and the RG equation can be combined, resulting in a
modified flow equation that is manifestly invariant under
global background rescalings.

\section{Introduction}

One of the most vexing challenges facing the asymptotic safety
approach to quantum gravity has been the double dependence
of the effective action on two fields, the background metric and
the fluctuation field.
It is only when both dependences are taken into account that
one can write an exact flow equation \cite{reuter1}.
On the other hand, physical results should be largely independent 
of the choice of background.
In fact, at the classical level, the action is invariant
under simultaneous transformations of the background and fluctuation.
At the simplest level, when one uses a linear parametrization
\be
g_{\mu\nu}=\bg_{\mu\nu}+h_{\mu\nu}
\label{linpar}
\ee
these are just the shift transformations
\be
\delta\bar g_{\mu\nu}=\epsilon_{\mu\nu}\ ,\qquad
\delta h_{\mu\nu}=-\epsilon_{\mu\nu} \ .
\ee
Ideally the effective action should also be invariant
under the same transformations.
However, the background gauge fixing procedure and the addition of
a cutoff term in the action spoil this invariance.
In much (in practice, up to 2008, all) work
on asymptotic safety this issue has been avoided by restricting
oneself to the so-called ``single field approximation''
where one sets the fluctuation field to zero.
The dangers of considering only the background dependence
had been pointed out in \cite{Litim:2002hj}.
In the last few years there have been several efforts
to address this issue.

One is to study bimetric truncations \cite{Manrique:2009uh}
and impose shift-invariance only in the IR limit \cite{Becker:2014qya}.
If one were able to calculate the whole bi-metric flow,
then the background flow could be obtained by setting the
classical fields to zero.
Thus in practice one method to improve on single-field truncations
is to keep as much as possible of the
fluctuation dependence, by calculating the flow of the
two-, three- and possibly four-point functions of the fluctuation
\cite{donkin,cdp,dep,pawlowski}.
Alternatively one can try to solve simultaneously the
Ward identity and the flow equation.
This could be achieved in the conformally reduced case
\cite{dm3,Labus:2016lkh,Dietz:2016gzg}.
Other related ideas have been discussed in \cite{Safari:2016gtj,Wetterich:2016ewc}.

A step forward has recently been made by Morris
for the special case when 
$\epsilon_{\mu\nu}=2\epsilon \bg_{\mu\nu}$,
{\it i.e.} when the background is simply rescaled by a constant factor
\cite{Morris:2016spn}.
He derived the modified Ward identity for this transformation
and showed that in six dimensions the anomalous terms
coming from the cutoff have the same form as the RG equation.
In this way the Ward identity and the RG equation can be
combined in a single equation that is amenable to
explicit treatment by the methods that are in current use.
The drawback of the proposed procedure is that
it only seems to work in six dimensions.

We show in this paper that modifying some steps of the procedure
is sufficient to obtain the same result in any dimension. 
The first and most crucial step is the replacement
of the linear split (\ref{linpar}) by the exponential
parametrization
\bea
g_{\mu\nu}= \bg_{\mu\rho}(e^\mX)^\rho{}_\nu\ ,
\label{exppar}
\ena
where
\be
X^\rho{}_\nu=\bg^{\rho\sigma}h_{\sigma\nu}\ .
\label{eugenio}
\ee
This parametrization is widely used in two-dimensional 
quantum gravity \cite{Kawai:1989yh}.
It has been introduceded in the functional RG setting in \cite{Eichhorn:2013xr,pv1}.
Its general virtues have been further discussed in \cite{nink,Codello:2014wfa,Gies:2015tca,pereiraI,pereiraII},
and it has been employed in several other explicit calculations
\cite{Eichhorn:2015bna,Labus:2015ska,opv,Dona:2015tnf}.

The second step is to make sure that no dimensionful parameter
enters the gauge-fixing term.
In the Einstein-Hilbert truncation it is convenient and customary
to have a prefactor $Z_N=1/(16\pi G)$,
so that the gauge-fixing terms combine smoothly with the Hessian,
but this introduces and unnecessary and, as we shall see, 
unwanted breaking of background scale invariance.
We will use a higher-derivative gauge-fixing,
which amounts to introducing some power of the Laplacian
in the gauge-fixing term.
This type of gauge fixing is often used with four-derivative
gravitational actions \cite{BS2,CP,cpr2,OP}
but normally not in the Einstein-Hilbert truncation.
There is however no fundamental reason for this,
other than simplicity \cite{pereiraII}.

The third step is to similarly avoid dimensionful
parameters in the cutoff term, except for the cutoff scale itself.
We will use a ``pure'' cutoff, namely one 
that does not contain any running parameter 
\cite{narain2,Labus:2015ska}.
As with the gauge-fixing term, 
in the Einstein-Hilbert truncation it is convenient 
to have a prefactor $1/(16\pi G)$.
In the $f(R)$ truncation the corresponding prefactor
is $-f'(R)$.
This dependence of the cutoff on running couplings
is however the source of unnecessary anomalies.

We will see that with these choices, the gauge-fixing becomes invariant
and the anomalous terms in the Ward identity
coming from the cutoff have the same form as the RG equation.
Then, the Ward identity expresses the invariance of the 
effective action under the transformation
of the background, fluctuation and a simultaneous
rescaling of the cutoff scale.
This identity can be solved and results simply in the
definition of new variables that are invariant under
background scale transformations.
The RG equation, written in these variables, 
no longer depends on the scale of the background metric 
and has the same form as the flow equation that is commonly used.
Although for the time being limited to simple scalings,
this points towards a practical solution of the
background-field dependence.

In section 2 we discuss the transformation of the fields
and of the gauge-fixing and cutoff actions.
In section 3 we derive the Ward identity and
combine it with the RG equation.
Section 4 contains a short discussion.

\section{Variations}

\subsection{Fields}

In this section we will often suppress indices and treat
two-index tensors as matrices.
Thus (\ref{exppar},\ref{eugenio}) will be written
$$
\mg=\mbg e^\mX\ ,\qquad \mX=\mbg^{-1}\mh\ .
$$
(Normally one would denote also $X$ by the symbol $h$,
but this would give rise to ambiguities when indices are suppressed.)
Note that $X$ is a linear map of the tangent space to itself,
so powers of $X$ and the trace of $X$ do not require use the
metric and are basis-independent.

Our first task is the following:
given an infinitesimal transformation $\delta\mbg$
of the background metric, find a transformation $\delta \mh$
of the fluctuation field such that the full metric $\mg$ is invariant.
We must have
\be
0=\delta \mg=\delta\mbg e^\mX+\mbg\delta e^\mX\ .
\label{invariance}
\ee
We use in (\ref{invariance}) the relation
\be
\delta e^\mX e^{-\mX}=\frac{e^{ad_\mX}-1}{ad_\mX}\delta \mX 
\label{carlo}
\ee
where $ad_\mX\mY=[\mX,\mY]$, and the relation 
$e^{\mX}\mY e^{-\mX}=e^{ad_\mX}\mY$,
to obtain
\be
\delta \mX=
-\frac{ad_\mX}{e^{ad_\mX}-1}
\mbg^{-1}\delta\mbg.
\label{laura}
\ee
Then, using Eq.~(\ref{eugenio}), we derive the variation of $\mh$:
\be
\delta \mh=\delta\mbg\mX+\mbg\delta\mX\ .
\label{alan}
\ee

Expanding
\be
\frac{ad_\mX}{e^{ad_\mX}-1}
=1-\frac{ad_\mX}{2}
+\frac{ad_\mX^2}{12}
-\frac{ad_\mX^4}{720}
+\frac{ad_\mX^6}{30240}
-\ldots
\ee
one could treat in this way general variations.
Things however simplify drastically when we consider
Weyl transformations
\be
\delta\bg_{\mu\nu}=2\epsilon \bg_{\mu\nu}\ ,
\label{shiftbg}
\ee
where $\epsilon$ is an infinitesimal transformation parameter
(a scalar function).
In this case $\mbg^{-1}\delta\mbg=2\epsilon\mathbf{1}$
is a multiple of the unit matrix, so
$\delta \mX=-2\epsilon\mathbf{1}$,
and thus using (\ref{alan})
\be
\delta h_{\mu\nu}=2\epsilon(h_{\mu\nu}-\bg_{\mu\nu})\ .
\label{shifth}
\ee
It is convenient to decompose the fluctuation field
into its tracefree and trace parts:
\be
h_{\mu\nu}=h^T_{\mu\nu}+\frac{1}{d} \bg_{\mu\nu} h
\label{decomp}
\ee
where $\bg^{\mu\nu}h^T_{\mu\nu}=0$.
We could further decompose the tracefree part
into spin-two, spin-one and spin-zero parts,
as in the York decomposition, but this is not necessary.
The following considerations hold whether one uses the York decomposition or not.

We have
\be
\delta h_{\mu\nu}=\delta h^T_{\mu\nu}
+\frac{1}{d} 2\epsilon \bg_{\mu\nu} h
+\frac{1}{d} \bg_{\mu\nu} \delta h\ .
\ee
On the other hand inserting (\ref{decomp}) in the r.h.s.
of (\ref{shifth}) and comparing the trace and tracefree parts
we find
\bea
\delta h^T_{\mu\nu}&=&2\epsilon h^T_{\mu\nu}\ ,
\nonumber\\
\delta h&=&-2d\epsilon\ .
\label{ennio}
\eea
Note that the tracefree
fluctuation transforms in the same way as the metric
whereas the trace transforms purely by a shift.
This is distinctly different from the behavior in the
linear decomposition (\ref{linpar})
and lies at the root of the subsequent simplifications.
In the special case when the manifold is compact and
$\epsilon$ is constant,
we can be even more specific.
If we decompose the trace into the constant part
and its orthogonal complement
\be
h=\underline h+\hperp\ ,
\ee
which is defined by the condition that its integral over
the whole manifold is zero,
the whole variation of $h$ is due to the constant component,
while $\hperp$ is invariant:
\be
\delta\underline h=-2d\epsilon\ ; \qquad
\delta\hperp=0\ .
\ee

We observe that if we restrict ourselves from the beginning to
Weyl transformations of the background metric,
there is a more direct derivation of (\ref{ennio}).
Raising one index in (\ref{decomp}) one can write
$\mX=\mX^T+\mathbf{1}h/d$, where $\mX^T$ is traceless.
Therefore
\be
\mg= \mbg\,  e^{h/d} e^{\mX^T}\ .
\label{alice}
\ee
If the background metric undergoes the finite transformation
$\mbg\to\mbg e^{2\epsilon}$, invariance of the full metric
can be maintained by the compensating tranformation
$h\to h-2d\epsilon$, while $\mX^T$ is left invariant.
Then $\delta\mh^T=\delta(\mbg \mX^T)=2\epsilon\mbg X^T=2\epsilon\mh^T$,
which is just (\ref{ennio}).
In the following we restrict ourselves to constant
Weyl transformations of the background metric.

\subsection{Gauge fixing}

Let us consider a gauge fixing term
\be
\label{gfaction}
S_{GF}=\frac{1}{2\alpha}\int d^d x \sqrt{\bg}\,F_\mu Y^{\mu\nu} F_\nu,
\ee
where $Y^{\mu\nu}$ is in general a differential operator,
\be
\label{gf}
F_\mu=\bnabla_\rho h^\rho{}_\mu-\frac{\beta+1}{d}\bnabla_\mu h
=\bnabla_\rho h^{T\rho}{}_\mu-\frac{\beta}{d}\bnabla_\mu h\ ,
\ee
and $\bnabla$ is the covariant derivative of $\bg_{\mu\nu}$.
Since the background Christoffel symbols are invariant
under background rescalings, taking into account also the variation
of the inverse metric that is hidden in $F_\mu$, one finds
\be
\delta F_\mu=0\ .
\ee
Let $\bDelta$ be a second-order Laplace-type operator constructed with
the background metric.
It tranforms under background rescalings as
\be
\delta\bDelta=-2\epsilon\bDelta\ .
\label{laptrans}
\ee
The gauge-fixing term will be invariant under background rescalings
if we choose
\be
Y^{\mu\nu}=\bDelta^{\frac{d-2}{2}}\bg^{\mu\nu}\ .
\ee

In order to derive the Faddeev-Popov operator, we start from
the transformation of the full 
metric under an infinitesimal diffeomorphism $\eta$,
which is given by the Lie derivative
\be
\label{transfg}
\delta_\eta g_{\mu\nu}=
\Lie_\eta g_{\mu\nu}
\equiv
\nabla_\mu\eta_\nu+\nabla_\nu\eta_\mu\ .
\ee
(Note that there are no bars on the $\nabla$s here.)
As usual, we have to define transformations of $\bg$ and $h$ that,
used in (\ref{exppar}), yield (\ref{transfg}).
The simplest one is the background transformation.
We use again matrix notation as in the preceding section.
If we treat $\mbg$ and $\mX$ as tensors under $\delta_\eta$, i.e.
\be
\delta^{(B)}_\eta\mbg=\Lie_\eta \mbg\ ;
\qquad
\delta^{(B)}_\eta \mX=\Lie_\eta\mX\ ,
\ee
then also
$\delta^{(B)}_\eta e^\mX=\Lie_\eta e^\mX$
and (\ref{transfg}) follows.
By definition, the ``quantum'' gauge transformation of 
$\mX$ is such as to reproduce (\ref{transfg})
when $\mbg$ is held fixed:
\be
\delta^{(Q)}_\eta \mbg=0\ ;
\qquad
\mbg\delta^{(Q)}_\eta e^\mX
=\Lie_\eta \mg
=\Lie_\eta\mbg e^\mX+\mbg\Lie_\eta e^\mX
\ .
\ee
From the latter relation we find
\be
e^{-\mX}\delta^{(Q)}_\eta e^\mX=
e^{-\mX}(\mbg^{-1}\Lie_\eta\mbg) e^\mX
+e^{-\mX}\Lie_\eta e^\mX
\ .
\ee
Then using (\ref{carlo}) one finds
\be
\delta^{(Q)}_\eta\mX=
\frac{ad_\mX}{e^{ad_\mX}-\mathbf{1}}
\left(
\mbg^{-1}\Lie_\eta\mbg
+\Lie_\eta e^\mX e^{-\mX}
\right)\ .
\ee

The Fadeev-Popov operator, acting on a ghost field $C_\mu$, 
is defined by
\be
\Delta_{FP\mu}{}^\nu C_\nu=
\bnabla_\rho\left(
(\delta^{(Q)}_C \mX)^\rho{}_\mu
+\frac{1+\beta}{d}\delta^\rho{}_\mu
\tr(\delta^{(Q)}_C \mX)
\right)
\label{elena}
\ee
where the infinitesimal transformation parameter $\eta$
has been replaced by the ghost.
The full ghost action then has the form \cite{pereiraII}
\be
S_{gh}(C^*_\mu,C_\mu;\bg_{\mu\nu})=\int d^dx\sqrt{\bg}\,
C^*_\mu Y^{\mu\nu}\Delta_{FP\nu}{}^\rho C_\rho \ .
\ee
Note that this action contains infinitely many interaction terms.
Expanding (\ref{elena}) to first order in $\mX$ we find:
\footnote{a factor $1/2$ is missing in equation (III.18) in \cite{pv1}.}
\be
\label{qtr}
\delta^{(Q)}_C \mX=
\mbg^{-1}\Lie_C\mbg
+\Lie_C \mX
+\frac{1}{2}[\mbg^{-1}\Lie_C\mbg,\mX]
+O(C \mX^2)\ .
\ee
In the single-metric truncation, where one puts $\mX=0$ from the start,
the Faddeev-Popov operator
is determined by the first term in this expansion.
It is a (generally non-minimal) Laplace-type operator
constructed with the background metric, and therefore 
transforms as in (\ref{laptrans}).
Invariance under global Weyl rescalings 
can be achieved simply demanding
\be
\delta C^\ast_\mu= 0\ ,\qquad
\delta C_\mu= 2\epsilon\, C_\mu\ .
\label{ghostscaling}
\ee
Then one can check that also the interaction terms are invariant.
An infinitesimal background rescaling
acting on $\Delta_{FP\mu}{}^\nu C_\nu$, as written in (\ref{elena}),
only affects on the terms $\delta^{(Q)}_C\mX$.
Since $\mX$ transforms by a constant shift,
$\delta (ad_\mX)=0$.
Then, because everything is linear in $C$,
$$
\delta\left(
\mbg^{-1}\Lie_C\mbg
+\Lie_C e^\mX e^{-\mX}
\right)
=-\epsilon\left(
\mbg^{-1}\Lie_C\mbg
+(\Lie_C e^\mX) e^{-\mX}
\right)
$$
and the remaining transformations involving $\mX$ cancel.
Notice the minus sign: this is due to the fact that the
Lie derivatives involve the contravariant field
$C^\mu=\bg^{\mu\nu}C_\nu$, whose transformation is $\delta C^\mu=-\epsilon C^\mu$.
So, finally
\be
\delta \Delta_{FP\mu}{}^\nu C_\nu
=-\epsilon\Delta_{FP\mu}{}^\nu C_\nu\ .
\ee
This, together with $\delta Y^{\mu\nu}=-d\epsilon Y^{\mu\nu}$
implies that the full ghost action is invariant.

The gauge-fixed action must also contain a term
\be
S_{aux}=\int dx\sqrt{\bg}B_\mu Y^{\mu\nu}B_\nu\ ,
\ee
where $B_\mu$ is an auxiliary bosonic field \cite{pereiraII}.
This Gaussian integral has the effect of removing the 
determinants of $Y$ from the effective action.
Scale-invariance is achieved provided the auxiliary field is inert:
$\delta B_\mu=0$.

We note that the procedure proposed here is by no means unique.
If one is interested mainly in the application of the formalism
to $f(R)$ theories \cite{cpr1,cpr2,benedetti,dm,dsz},
where one normally considers a spherical background,
then one could define $Y^{\mu\nu}=\bR^{\frac{d-2}{2}}\bg^{\mu\nu}$.
This achieves scale invariance without having to introduce
an auxiliary field, but it would not work on a flat background.
One could also have a mix of $\bDelta$ and $\bR$,
provided the overall power is $\frac{d-2}{2}$.
Yet another choice would be the ``physical gauge''
advocated in \cite{pv1}.
In this case one would just set $\hperp=0$ and $\xi_\mu=0$,
where $\xi_\mu$ is the spin-one degree of freedom of $h_{\mu\nu}$.
Since $\hperp$ is invariant and $\xi_\mu$ trasforms homogeneously
under scaling, these conditions are scale-invariant.
They produce Faddeev-Popov determinants that can
be taken care of by introducing suitable auxiliary fields.

\subsection{Cutoff term}

Next we consider the cutoff term, which lies at the root of the issue.
It has the general structure
\be
\Delta S_k(h_{\mu\nu};\bg_{\mu\nu})=
\frac{1}{2}\int d^dx\sqrt{\bg}\,
h_{\mu\nu}\cR_k^{\mu\nu\rho\sigma}
h_{\rho\sigma}\ ,
\label{gencutoff}
\ee
where $\cR_k^{\mu\nu\rho\sigma}(\bDelta)$, in coordinate space,
is a two-point kernel.
It is typically chosen to have the form
\be
\cR_k^{\mu\nu\rho\sigma}(\bDelta)=
\frac{1}{2}\left(\bg^{\mu\rho}\bg^{\nu\sigma}
+\bg^{\mu\sigma}\bg^{\nu\rho}
+a\bg^{\mu\nu}\bg^{\rho\sigma}\right)
c \,k^{d-2} R_k(\bDelta)
\ee
where $a$ and $c$ are dimensionless constants
and $R_k(0)=k^2$, with $k$ the IR cutoff scale which controls the coarse-graining procedure. Usually one defines the RG "time" as $t \sim \ln k$.
By dimensional analysis 
\be
R_k(\bDelta)=k^2 r(y)\ , \qquad y=\bDelta/k^2\ ,
\label{diman}
\ee
where $r$ is a dimensionless function that goes rapidly to zero
for $y>1$ and $r(0)=1$.

In the Einstein-Hilbert truncation and in de Donder gauge 
it is very convenient to choose $a=-1$, so that the tensor
structure matches the one of the Hessian 
(including the gauge-fixing term). 
Furthermore, it is almost always assumed that 
$c\, k^{d-2}=1/(16\pi G)$.
Then the cutoff combines seamlessly with the Hessian
resulting simply in the substitution of
$\bDelta\to\bDelta+R_k(\bDelta)$, where 
in this specific case $\bDelta=-\bnabla^2$.
We are not committed to using any specific form of the action here,
so we leave the constants $a$ and $c$ unspecified.
Such a cutoff is then called ``pure'' to emphasize that it
does not contain any running coupling.

Similarly one introduces the cutoff operator for the ghosts.
Using the decomposition (\ref{decomp}) we can write
\bea
\Delta S_k(h^T_{\mu\nu},h;\bg_{\mu\nu})\!\!\!&=&\!\!\!
\frac{1}{2}\int d^dx\sqrt{\bg}\,
\left[
h^T_{\mu\nu}\bg^{\mu\rho}\bg^{\nu\sigma}\cR^T_k(\bDelta)
h^T_{\rho\sigma}
+h \cR_k(\bDelta) h\right]\ ,
\label{2dcutoff}
\\
\Delta S^{gh}_k(C^*_\mu,C_\mu;\bg_{\mu\nu})\!\!\!&=&\!\!\!
\int d^dx\sqrt{\bg}\,
C_\mu^* \bg^{\mu\nu}\cR_k^{gh}(\bDelta) C_\nu\ ,
\\
\Delta S^{aux}_k(B_\mu;\bg_{\mu\nu})\!\!\!&=&\!\!\!
\int d^dx\sqrt{\bg}\,
B_\mu^* \bg^{\mu\nu}\cR_k^{aux}(\bDelta) B_\nu\ ,
\eea
where $\cR_k^T=c\, k^{d-2}R_k=c\,k^d r(y)$, 
$\cR_k=c_0 k^{d-2}R_k=c_0 k^d r(y)$ 
with $c_0=c\frac{2+ad}{2d}$,
$\cR_k^{gh}(\bDelta)=c_{gh}k^{d-2}R_k(\bDelta)=c_{gh} k^d r(y)$
and
$\cR_k^{aux}(\bDelta)=c_{aux}k^{d-4}R_k(\bDelta)=c_{aux}k^{d-2} r(y)$.

The Laplacian transforms under background rescalings as
in (\ref{laptrans}).
Since $k$ does not change under a variation of the background metric,
we find from (\ref{diman}) $\delta \cR_k=-2\epsilon k^d y r'$.
On the other hand $\partial_t \cR_k=d k^d r-2k^d y r'$,
so
\be
\delta \cR_k=\epsilon(-d\cR_k+\partial_t \cR_k)\ .
\label{piero}
\ee
The kernel $\cR_k^T$ transforms in the same way.
The remarkable fact is that the first term on the r.h.s.
exactly cancels the variation of the volume element.
Since the tracefree fluctuation transforms homogeneously,
in the same way as the covariant metric,
the variations of the inverse metric and those of the
fields $h^T_{\mu\nu}$ also cancel.
Thus the variation of (\ref{2dcutoff}) is
\be
\delta\Delta S_k(h^T_{\mu\nu},h;\bg_{\mu\nu})=
\frac{1}{2}\epsilon \int d^dx\sqrt{\bg}
\left[
h^T_{\mu\nu}\bg^{\mu\rho}\bg^{\nu\sigma}\partial_t \cR_k^T
h^T_{\rho\sigma}
+h\partial_t\cR_k h\right]
-2d\epsilon\int d^dx\sqrt{\bg}\,\cR_k h\ ,
\label{varcuth}
\ee
where the last term comes from the variation of the 
trace fluctuation $h$.

The ghost cutoff kernel $\cR_k^{gh}$ also transforms as in 
(\ref{piero}) so that
\be
\delta\Delta S_k^{gh}(C^*_\mu,C_\mu;\bg_{\mu\nu})=\epsilon \int d^dx\sqrt{\bg}\,
C_\mu^*\bg^{\mu\nu}\partial_t\cR_k^{gh} C_\nu\ .
\label{varcutgh}
\ee

Finally, the variation of the auxiliary term works a bit differently.
Instead of (\ref{piero}) one has
\be
\delta \cR_k^{aux}=
\epsilon(-(d-2)\cR_k^{aux}+\partial_t \cR^{aux}_k)\ .
\label{piera}
\ee
The first term exactly cancels the transformation due to the measure
and inverse metric, so that again
\be
\delta\Delta S_k^{aux}(B_\mu;\bg_{\mu\nu})=\epsilon 
\int d^dx\sqrt{\bg}\,
B_\mu\bg^{\mu\nu}\partial_t\cR_k^{aux} B_\nu\ .
\label{varcutaux}
\ee

We note that in comparison with \cite{Morris:2016spn}
all the terms proportional to $\Delta S_k$ itself,
that came with a factor $d-6$, are absent here.

\section{The Ward identity}

We now have all the ingredients that are needed to derive the
Ward identity.
The Effective Average Action (EAA) is defined by
\bea
\Gamma_k(\bh^T_{\mu\nu},\bar h,\bar C^*_\mu,\bar C_\mu,\bar B_\mu;\bg_{\mu\nu})
&=&-W_k(j_T^{\mu\nu},j,\jgh^\mu_*,\jgh^\mu,\jaux^\mu;\bg_{\mu\nu})
\nonumber\\
&&
+\int d^dx \left(j_T^{\mu\nu}\bar{h}^T_{\mu\nu}+j\bar h
+\jgh_*^\mu \bar C^*_\mu+\jgh^\mu \bar C_\mu
+\jaux^\mu \bar B_\mu\right)
\nonumber\\
&&
-\Delta S_k(\bar{h}^T_{\mu\nu},\bar h;\bg_{\mu\nu})
-\Delta S_k^{gh}(\bar C^*_\mu,\bar C_\mu;\bg_{\mu\nu})
-\Delta S_k^{aux}(\bar B_\mu;\bg_{\mu\nu})\ ,
\nonumber
\label{eaa}
\eea
where $W_k$ is the generating
functional of connected Green functions,
$\bar{h}^T_{\mu\nu}$, $\bar h$ etc. denote here
the classical VEVs of the corresponding quantum fields, 
the sources $j_T^{\mu\nu}$, $j$, $\jgh_*^\mu$, $\jgh^\mu$
and $\jaux^\mu$ have to be interpreted as usual
as functionals of these classical fields
and the last three term subtracts the cutoff that had been
added in the beginning to the bare action. 

The modified Ward identity for $\Gamma_k$ can be obtained as in \cite{Morris:2016spn}
by first varying $W_k$ and then using the Legendre transform.
Alternatively, we can start from
the integro-differential functional equation 
\bea
\label{ea_equation}
e^{-\Gamma_k(\bh^T_{\mu\nu},\bar h,\bar C^*_\mu,\bar C_\mu,\bar B_\mu;\bg_{\mu\nu})}\!\!\!\!
&=& \!\!\!\!
\int \!\! D h^TDh DC^* DC DB
\, Exp[-S-S_{GF}-S_{gh}-S_{aux}]
\\
&& \hspace{-4cm}
\times\, Exp\int\left[ 
\frac{\delta\Gamma_k}{\delta\bh^T} 
(h^T-\bh^T)
+\frac{\delta\Gamma_k}{\delta\bh} (h-\bh)
+\frac{\delta\Gamma_k}{\delta C}(C-\bar{C})
+\frac{\delta\Gamma_k}{\delta \bar{C}^\ast}(C^\ast-\bar{C^\ast})
+\frac{\delta\Gamma_k}{\delta B}(B-\bar{B})
\right]
\nonumber\\
&&\hspace{-4cm}
\times\, Exp\left[
-\Delta S_k(h^T-\bh^T,h-\bh;\bg)
-\Delta S^{gh}_k(C^*-\bar C^*,C-\bar C;\bg)
-\Delta S_k^{aux}(B-\bar B;\bg)
\right]
\nonumber
\eea
where a bar over a field denotes its vacuum expectation value
and we have suppressed all indices for typographycal clarity.
Varying both sides
\bea
\delta \Gamma_k &=& 
-\int \frac{\delta\Gamma_k}{\delta \bar{h}^T} 
\langle \delta h^T-\delta\bh^T\rangle
-\int \frac{\delta\Gamma_k}{\delta\bh} 
\langle\delta h-\delta \bh\rangle
\\
&&
-\int \frac{\delta\Gamma_k}{\delta \bar C} 
\langle\delta C-\delta \bar C\rangle
-\int \frac{\delta\Gamma_k}{\delta \bar C^\ast} 
\langle\delta C^\ast-\delta \bar C^\ast\rangle
-\int \frac{\delta\Gamma_k}{\delta \bar B} 
\langle\delta B-\delta \bar B\rangle
\nonumber\\
&&
+\langle \delta\Delta S_k(h^T-\bh^T,h-\bh;\bg)\rangle
+\langle\delta\Delta S^{gh}_k(C-\bar C;\bg)\rangle
+\langle\Delta S_k^{aux}(B-\bar B;\bg)\rangle
\ .
\nonumber
\eea

The variations (\ref{ennio}) and those of the ghost and auxiliary fields
are at most linear in the fields.
Thus $\langle\delta\phi\rangle=\delta \bar\phi$ for all fields.
All the terms in the first two lines are therefore zero and
the only anomalous contribution comes from the 
cutoff terms.
Using the variation in Eq.~(\ref{varcuth},\ref{varcutgh},\ref{varcutaux}) 
one finds
\bea
\hspace{-1cm}\delta \Gamma_k &=& 
\langle \delta\Delta S_k(h^T-\bh^T,h-\bh;\bg)\rangle
+\langle\delta\Delta S^{gh}_k(C-\bar C;\bg)\rangle
+\langle\delta\Delta S^{aux}_k(B-\bar B;\bg)\rangle
\nonumber\\
\label{wir}
&=&
\epsilon\Bigg[ 
\frac{1}{2}\Tr\left(
\frac{\delta^2\Gamma_k}{\delta \bar{h}^T\delta \bar{h}^T}
+\cR_k^T\right)^{-1}\!\!\!\partial_t \cR_k^T
+\frac{1}{2}
\Tr\left(
\frac{\delta^2\Gamma_k}{\delta \bar{h}\delta \bar{h}}+\cR_k\right)^{-1}\!\!\!\partial_t \cR_k
\\
&&\qquad
-\Tr\left(
\frac{\delta^2\Gamma_k}{\delta \bar{C^\ast}\delta \bar{C}}+\cR_k^{gh}\right)^{-1}\!\!\!\partial_t\cR_k^{gh}
+\frac{1}{2}
\Tr\left(
\frac{\delta^2\Gamma_k}{\delta \bar B\delta \bar B}+\cR^{aux}_k\right)^{-1}\!\!\!\partial_t \cR^{aux}_k+\ldots
\Bigg]\,.\nonumber
\eea
Apart from the factor $\epsilon$,
the r.h.s. is identical to the r.h.s. of the exact RG equation.
(The ellipses stand for terms involving mixed 
functional derivatives that are 
present in the exact equation but are neglected in common approximations.)
The l.h.s. of the identity is the total variation of $\Gamma_k$,
with $k$ held fixed, which can be expressed as:
\be
\delta\Gamma_k=
\epsilon\int d^dx\left[
2\bg_{\mu\nu}\frac{\delta\Gamma_k}{\delta \bg_{\mu\nu}}
+2 h^T_{\mu\nu}\frac{\delta\Gamma_k}{\delta h^T_{\mu\nu}}
-2d\frac{\delta\Gamma_k}{\delta h}
+2  C_\mu\frac{\delta\Gamma_k}{\delta C_\mu}
\right]\ .
\label{wil}
\ee
Here and in the following we omit 
the bars over $h^T_{\mu\nu}$ and $h$,
since the argument of $\Gamma_k$ are always the classical
expectation values and no confusion can arise.

\section{The Ward identity and the flow equation}

We have arrived at a remarkably simple result:
the anomalous variation in the background scale Ward identity
is exactly the ``beta functional'' of the theory induced by the coarse-graining procedure, as expressed
by the r.h.s. of the RG equation:
\be
\delta_\epsilon\Gamma_k=\epsilon\,\partial_t\Gamma_k\ ,
\label{main}
\ee
where we recall that the variation on the l.h.s.
involves only the functional arguments of $\Gamma_k$
and leaves $k$ fixed.
Bringing the r.h.s. to the l.h.s. we obtain that
\be
\int d^dx\left[
2\bg_{\mu\nu}\frac{\delta\Gamma_k}{\delta \bg_{\mu\nu}}
+2 h^T_{\mu\nu}\frac{\delta\Gamma_k}{\delta h^T_{\mu\nu}}
-2d\frac{\delta\Gamma_k}{\delta h}
+2 C_\mu\frac{\delta\Gamma_k}{\delta C_\mu}
\right]
-k\frac{d\Gamma_k}{dk}=0\ .
\label{wig}
\ee
This is just the statement that the EAA is invariant
under scalings of the background metric, accompanied by
suitable transformations of the other fields
{\it and} by a rescaling of the cutoff $k$:
\be
\delta k=-\epsilon k\ .
\ee
As discussed in \cite{Morris:2016spn}, (\ref{wig})
can be solved using the method of characteristics.
One must have
\be
\frac{d\bg_{\mu\nu}}{d\lambda}=2\bg_{\mu\nu}
\ ;\quad
\frac{d h^T_{\mu\nu}}{d\lambda}=2 h^T_{\mu\nu}
\ ;\quad
\frac{d \underline{h}}{d\lambda}=-2d
\ ;\quad
\frac{d C_\mu}{d\lambda}=2C_\mu
\ ;\quad
\frac{d k}{d\lambda}=-k\ ,
\ee
whose solutions are simply
\bea
&&\bg_{\mu\nu}(\lambda)=e^{2\lambda}\bg_{\mu\nu}(0)
\ ;\quad
h^T_{\mu\nu}(\lambda)=e^{2\lambda}h^T_{\mu\nu}(0)
\ ;\quad
\underline{h}(\lambda)=\underline{h}(0)-2d\,\lambda
\ ;
\nonumber\\
&&
C_\mu(\lambda)=e^{2\lambda}C_\mu(0)
\ ;\quad
k(\lambda)=e^{-\lambda} k(0)\ ,
\eea
while $\hperp$, $C^*_\mu$ and $B_\mu$ are constant.
The last relation implies that $\lambda=-t$.
Thus the scaling parameter can be identified with the RG time.
The combinations
\be
\hat k=e^{-\underline{h}/2d}k\ ;\qquad
\hat g_{\mu\nu}=e^{\underline{h}/d}\bg_{\mu\nu}\ ;\qquad
\hat{h}^T_{\mu\nu}=e^{\underline{h}/d}h^T_{\mu\nu}\ ;\qquad 
h^\perp\ ;\qquad
\hat C_\mu=e^{\underline{h}/d}C_\mu
\ee
are invariant.~\footnote{Alternatively one could also define
$\hat k=\bar V^{1/d}k$, where $\bar V$ is the volume in the metric $\bg_{\mu\nu}$.}
The solution of the Ward identity is therefore a functional 
\be
\Gamma_k(h^T_{\mu\nu},h^\perp,\underline{h},C^*_\mu,C_\mu,B_\mu;\bg_{\mu\nu})
=\hat\Gamma_{\hat k}(\hat{h}^T_{\mu\nu},h^\perp,C^*_\mu,\hat C_\mu,B_\mu;\hat{g}_{\mu\nu})\ .
\ee
As expected the Ward identity eliminates the dependence of
the EAA on the variable $\underline h$ and on the total volume
of the background metric, replacing it by the dependence
on the total volume of the metric $\hat g_{\mu\nu}$.
The solution can be written entirely in terms of quantities
that are invariant  under constant Weyl rescaling of the background.
In particular, if one specializes to the case when 
$h^\perp=0$, $h^T_{\mu\nu}=0$, $C^*_\mu=0$, $C_\mu=0$, $B_\mu=0$
one has
\be
\Gamma_k(\underline{h};\bg_{\mu\nu})=
\hat\Gamma_{\hat k}(\hat{g}_{\mu\nu})\ .
\ee
Note that if we set $h^T_{\mu\nu}=0$ and $\hperp=0$, 
then $\hat{g}_{\mu\nu}$
is the classical value of the full quantum metric, see (\ref{alice}).

If we were able to solve the full Ward identities 
related to arbitrary deformations of the background,
we would obtain a functional $\hat\Gamma_{\hat k}(\hat{g}_{\mu\nu})$,
that would satisfy a flow equation containing its
second derivatives with respect to $\hat g_{\mu\nu}$.
Having only partly transferred the field dependence 
from the fluctuation field to $\hat g_{\mu\nu}$,
we will have a flow equation containing second derivatives
with respect to the remaining fluctuation fields
{\it and} second derivatives with respect to those 
deformations of $\hat g_{\mu\nu}$ that have become dynamical
as a result of solving the Ward identity.
(In the present case, this is just the overall scale
of $\hat g_{\mu\nu}$.)
This distinction obviously gets blurred when one uses the 
single-metric approximation.

\section{Discussion}

The main outcome of this paper is the generalization of the
results of \cite{Morris:2016spn} for the background scale
Ward identity in quantum gravity.
Morris was able to show that in six dimensions 
the violation of background scale invariance is given exactly 
by the r.h.s. of the RG equation.
This is reminiscent of the statement that in a classically
scale-invariant quantum field theory in flat space, 
such as masseless QCD,
the violation of scale invariance is proportional to the beta functions.
The physical meaning of the identity is different in the two cases:
in QCD it is a genuine anomaly, whereas in quantum gravity the
anomalous variation under a change of background can be absorbed
by a change of the fluctuation field $\underline h$
{\it and} of the cutoff $k$, as we have seen in the preceding section.
Nevertheless, the two statements are formally the same, and
one would expect such general statements to be true
in any dimension.
Indeed we have shown here that this is the case.

To get this result, however, one has to make certain
choices that minimize the breaking of scale invariance.
The main difference with \cite{Morris:2016spn} is the use of the
exponential parametrization for the metric (\ref{exppar}).
When the linear split (\ref{linpar}) is used,
invariance of the full metric requires that the fluctuation field
has a transformation opposite to the one of the background field,
with the exception of the trace that has a mixed transformation
consisting of a homogeneous and an inhomogeneous term.
With the exponential  parametrization, the fluctuation field
transforms in {\it the same} way as the background metric,
with the exception of the trace that transforms purely
by a shift, in much the same way as a dilaton.
These transformation rules merely reflect the dimensions
of the fields (when the coordinates are dimensionless and
a metric has given dimension of area)
and the remaining choices also follow the dimensions
of each field.
The other differences are in the gauge-fixing and cutoff terms:
one has to make sure that these do not contain dimensionful
couplings that would introduce {\it additional}
unwanted scale-breaking terms.
Of course, it is unavoidable to break background scale invariance
by introducing the cutoff scale $k$, but the main point of the
present exercise has been to show that if this is the {\it only}
source of scale-breaking, its effect is entirely contained in
the RG flow of the couplings.
To this effect, in $d>2$ we have used a higher-derivative cutoff,
such as is used in higher-derivative gravity,
and a ``pure'' cutoff, that does not contain any Lagrangian parameter.
We stress that with this gauge fixing we are able 
to prove invariance of the ghost action including
all ghost interactions.
There may be other procedures that also work well,
but these three choices are sufficient to ensure that
the Ward identity does not contain additional, 
unnecessary anomalous terms.

The Ward identity can be used 
to reduce the number of variables
that the effective average action depends upon.
Ultimately one would like to reduce the flow equation for
$\Gamma_k(h_{\mu\nu};\bg_{\mu\nu})$
to a flow equation for a functional of a single field
$\hat\Gamma_{\hat k}(\hat g_{\mu\nu})$.
Reference \cite{Morris:2016spn} and the present work
are first steps towards background independence:
we have shown how to eliminate from the RG flow 
the dependence on a single
real degree of freedom: the overall scale of the background.
This may look like a rather small step,
but without it the beta functions are likely to contain spurious terms.
We plan to investigate this in concrete calculations.
The main value of the present work may lie in
restricting the freedom of choice of parametrization,
gauge and cutoff scheme.
Eq.~(\ref{main}) is an important statement, 
even if restricted to constant Weyl transformations:
it is expected of any quantum field theory that
is invariant under rescalings of the background metric
at the classical level.
One should be wary of using parametrizations and/or cutoff schemes
that violate it.

\section*{Acknowledgment}
G.P.V thanks Jan M. Pawlowski for interesting discussions.


\end{document}